# A SUPERCONDUCTIVE, LOW BETA SINGLE GAP CAVITY FOR A HIGH INTENSITY PROTON LINAC

A. Facco, V. Zviagintsev[1], INFN, Laboratori Nazionali di Legnaro, Italy
B. M. Pasini, Universita' di Padova, Italy

*Abstract*

A single gap, 352 MHz superconducting reentrant cavity for 5-100 MeV beams has been designed and it is presently under construction. This development is being done in the framework of a 30 mA proton linac project for nuclear waste transmutation. Mechanical, cryogenic and rf design characteristics of such cavities will be described.

## 1 INTRODUCTION

The TRASCO project for nuclear waste transmutation [1] requires a 5-100 MeV linac for acceleration of a 30 mA proton beam. Generally, room temperature Drift Tube Linac structures are used in this energy range; however, since the high duty cycle required for high current beams implies a very high power density on the resonators walls, the superconducting solution would offer many advantages. Among various possibilities, single gap superconducting cavities at 352 MHz, able to cover the full range of velocity, could allow a linac design with high reliability [2].

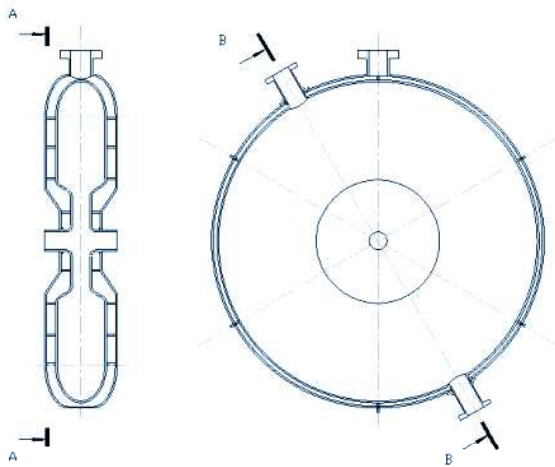

Fig.1 Mechanical design of the reentrant cavity.

This Independently-phased Superconducting Cavities Linac (ISCL) scheme, in which the power is distributed in a large number of rf sources, allows to increase the expected beam MTBF (Mean Time Between Failures) significantly [3]; this is of outmost importance for ADS applications. Keeping the beam loading within about 15 kW per cavity allows beam transport without losses even in case of failure of one of the cavities, with no need to reset immediately the linac parameters. This property gives the time for on-line repair or substitution of accessible failing components, like rf amplifiers and power supplies, which are usually the main source of beam shutdown [4]. We have designed and we are building a single gap cavity that could fulfil the required characteristics for an ISCL, i.e.: a relatively large bore to reduce the risk of beam losses; axial symmetry of the accelerating field to prevent emittance increase (e.g., short quarter wave resonators usually have a measurable dipole field component); very wide velocity acceptance given by the single gap, allowing for only one resonator type along the whole linac; simple geometry, which gives the possibility of optimising the production cycle for a low cost. A further advantage of this design is the possibility of using solid state rf amplifiers, which can be presently built at a competitive cost but with an expected increase of lifetime and reliability compared to tetrode and IOT amplifiers [9].

In the past, the main difficulty related to the construction of pillbox type superconducting cavities was the strong multipacting. Reentrant cavities with tolerable multipacting have been demonstrated at Stanford [5]. Another difficulty related to axially symmetric low-$\beta$ cavities is their flat shape, which requires a very strong mechanical structure to hold the liquid helium pressure.

## 2 CAVITY DESIGN

Since our peculiar constraint was the linac reliability, we designed this cavity with the conservative assumptions of operating it at a maximum surface field of 25 MV/m, with a total rf surface resistance up to 100 $n\Omega$ ($R_{BCS} \cong 58 n\Omega$ at 4.2 K) ; according to the linac architecture [8], a maximum beam loading of 15 kW is foreseen. The study included rf design, multipacting simulations and mechanical design.

---
[1] on leave from ITEP, Moscow, Russia

The rf design was studied with the code SUPERFISH; the calculated cavity parameters are listed in tab. 1. The resonator, with an rf length of 80 mm and a total length of 135 mm along the beam axis, is rather short and allows for a good linac packing factor; its outer diameter is 542 mm. Both the bore diameter and the gap length are 30 mm, allowing efficient acceleration down to β=0.1, i.e. about 5 MeV/A. The design value of the accelerating field in operation is 8.3 MV/m, requiring less than 7W per cavity. We should remind that the search for maximum shunt impedance was not the aim of this work. Since the main limitation in low-β superconducting cavities is usually determined by the maximum achievable surface electric field, related to the onset of field emission, and since our linac design required relatively low energy gain per cavity, we looked mainly for low surface electric and magnetic field, and short size.

| Total length | 135 | mm |
|---|---|---|
| Internal length | 80 | mm |
| bore radius | 15 | mm |
| frequency | 352 | MHz |
| $U/E_a^2$ | 0.034 | J/(MV/m)$^2$ |
| $E_p/E_a$ | 3.05 | |
| $H_p/E_a$ | 30.6 | Gauss/(MV/m) |
| $\Gamma = R \times Q$ | 83.9 | $\Omega$ |
| $R'_{sh}$ (Cu) | 18 | M$\Omega$/m |
| β | $\geq 0.1$ | |

Tab. 1 Resonator parameters calculated by means of the code SUPERFISH.

The rf power will be fed through a inductive coupler, still under study, located on the resonator equator; a coaxial capacitive coupler in the resonator cut-off tube, although preferable, would be hardly feasible, due to the constraints on the resonator length imposed by the required linac packing factor.

A deep study of the multipacting (MP) properties of the cavity was performed, with more than 50000 runs of the simulation program TWTRAJ [6]. The shape of the gap, where strong MP could be expected, prevents the onset of MP because of the absence of the dangerous values of the electric-to-magnetic field ratio. The region where most of the multipacting levels were concentrated, in the initial design schemes, was the resonator equator. Electrons originated in many different positions inside the resonator and drifting to the equator were collected there (see Fig. 2), building very strong levels with various multiplicities. This region was properly shaped with an elliptical contour (ratio between axes 1:1.5) and all high field levels have been finally eliminated.

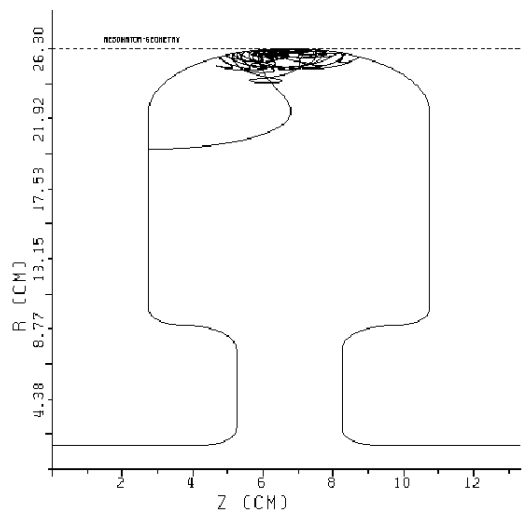

Fig.2. Example of the multipacting simulation program output (the horizontal axis is stretched in the picture). The circular contour, as well as the rectangular one, was causing high field multipacting levels.

The main concern in the mechanical design was related to the large force acting on the relatively flat and wide resonator surface. The niobium sheet, 3 mm thick, could not sustain the pressure in the absence of strong reinforcement. We tackled this problem using a different approach: in our design, the helium vessel is part of the resonator (like in the LNL bulk niobium low beta cavities [10]), and it is welded to the resonator wall so that the net force on the total structure is nearly cancelled. This design allows a relatively light structure with a good stability and a minimum displacement of the walls under pressure (limited to a few microns) when the tuner is mounted. The mechanical stresses are also confined to rather safe values (Fig. 3,4). The 4.2K helium is fed by gravity through a flange on the top of the resonator; like in most superconducting linacs for heavy ions, there is no separation between the beam vacuum and the cryostat one.

The tuning is obtained, changing the gap length, by means of a mechanical tuner connected to the external part of the drift tubes. A "nutcracker" type tuner, driven by a piezoelectric or magnetostrictive actuator, is being presently considered. Since the required tuning force is relatively low, the calculated Lorenz force detuning ($\cong 170$ Hz/(MV/m)$^2$) and helium pressure detuning ($\cong 140$ Hz/mbar) of the bare cavity are not negligible in comparison to the 1 kHz resonator rf bandwidth in operation. A stiff tuner reduces the pressure detuning to $\cong 20$ Hz/mbar. The foreseen cw mode of operation gives a nearly constant radiation pressure on the walls that can be compensated by the tuner. The experimental study of possible mechanical instabilities and of their remedies, however, is one of the aims of this prototype construction. For the design and simulation of the mechanical structure we used the I-DEAS code [7].

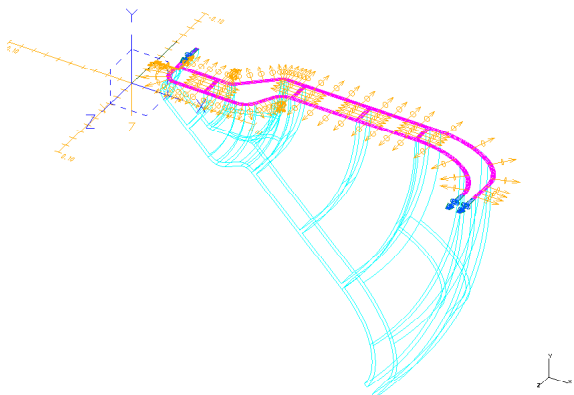

Fig.3. Reentrant cavity 2D mechanical simulation.

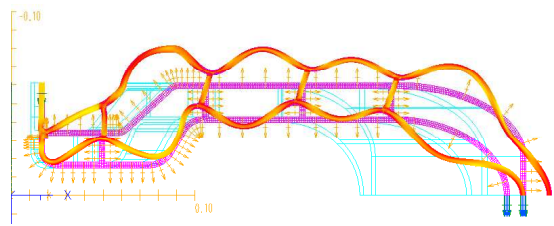

Fig. 4. Mechanical deformation as a result of 1 bar pressure increase in the helium bath. The maximum displacement is about 20 µm in the outer shell (for clarity, the deformation is amplified in the picture).

## 3  CONCLUSIONS

A low beta, 352 MHz superconducting reentrant cavity for high intensity proton beams has been designed. This single-gap structure contains novel features in the mechanical construction and allows for acceleration of low beta, 5-100 MeV protons. The cavity was designed as the accelerating element for an Independently-phased Superconducting Linac (ISCL), characterized by high reliability and efficiency, for ADS applications. The prototype construction has started and the first rf test is expected within 2000.

## ACKNOWLEDGEMENTS

We are grateful to M. Comunian and A. Pisent for their collaboration and their clarifying suggestions. We thank also H. Safa and R. Parodi for the fruitful discussions about multipacting.